\documentclass[twocolumn,showpacs,preprintnumbers,amsmath,amssymb]{revtex4}

\usepackage{graphicx}
\usepackage{dcolumn}
\usepackage{bm}

\begin{document}


\title{Consistent reasoning about a continuum of hypotheses\protect\\ on the basis of finite evidence}

\author{Jochen Rau}
 \email{jochen.rau@web.de}
\affiliation{%
Rh\"{o}nstra{\ss}e 70, 60385 Frankfurt, Germany
}%


\date{\today}

\begin{abstract}
In the modern Bayesian view classical probability theory is simply an extension of conventional logic, i.e., a quantitative tool that allows for consistent reasoning in the presence of uncertainty.
Classical theory presupposes, however, that---at least in principle---the amount of evidence that an experimenter can accumulate always matches the size of the hypothesis space.
I investigate how the framework for consistent reasoning must be modified in non-classical situations where hypotheses form a continuum, yet the maximum evidence accessible through experiment is not allowed to exceed some finite upper bound.
Invoking basic consistency requirements pertaining to the preparation and composition of systems, as well as to the continuity of probabilities, I show that the modified theory must have an internal symmetry isomorphic to the unitary group.
It thus appears that the only consistent algorithm for plausible reasoning about a continuum of hypotheses on the basis of finite evidence is furnished by quantum theory in complex Hilbert space.
\end{abstract}

\maketitle

\section{\label{introduction}Introduction}

In the modern Bayesian view classical probability theory with its two key ingredients (i) Bayes' learning rule and (ii) maximum entropy priors, is nothing but an extension of conventional logic, i.e., a quantitative tool that allows for consistent  reasoning also in the presence of uncertainty \cite{jaynes:book,sivia:book}.
Probabilities are no longer defined as limits of relative frequencies but as ``degrees of belief'' that are subject to certain consistency requirements \cite{cox:probability}, and that can be legitimately assigned not just to ensembles but also to individual systems.
Bayesian probability theory is thus more broadly applicable than the orthodox frequentist approach, while yielding identical results in those cases where a large $N$ limit exists.

Quantum theory is inherently probabilistic: 
Does it therefore, too, lend itself to a Bayesian interpretation \cite{caves:quantumasbayes,srednicki:probabilities}?
More specifically, does quantum theory, too, represent some kind of ``optimal algorithm'' for plausible reasoning in a certain, yet to be specified setting?
There are indications that this might be the case, as quantum theory has been linked to concepts such as a modified propositional calculus, ``learning'', and---most recently---information processing:
(i) One of the earliest attempts (long before the advent of modern Bayesianism) to axiomatize quantum theory started from a generalisation of classical propositional calculus, relaxing the requirement that all propositions be jointly decidable and resulting in a mathematical structure dubbed ``quantum logic'' \cite{birkhoff:logic,jauch:book,varadarajan:book,cohen:book};
the key result of this approach being that propositions within (an irreducible building block of) such a ``quantum logic'' can always be identified with subspaces of a Hilbert space over some skew field \cite{piron:axiomatique}.
(This approach fails, however, to give a compelling argument why the skew field should be the complex numbers, and does not work for dimension two.)
(ii) The discontinuous change of the density matrix upon quantum measurement has been shown to be closely related to Bayes' learning rule \cite{schack:bayesrule}.
(iii) Ongoing research in the fast-growing field of quantum information and quantum computation keeps revealing intimate connections of quantum theory with, and its potential power for, information processing \cite{nielsen:book,steane:computing,peres:rmp,keyl:fundamentals}.
There is even a recent proposal to reduce the key features of quantum theory---albeit not the full Hilbert space structure---to a small number of purely information-theoretic constraints \cite{clifton:constraints}.

In  this paper I attempt to pinpoint the circumstances under which, and the sense in which, the basic laws of quantum theory may indeed be considered an ``optimal'' set of rules to conduct plausible reasoning in the presence of uncertainty.
But how is this possible if classical Bayesian theory is already thought to be \textit{the} universal algorithm?
The basic idea is the following.
In a probabilistic model every proposition can be built up, through logical operations, from a certain minimal set---called the ``hypothesis space''---of elementary propositions. 
Classical probability theory assumes that all these elementary propositions are jointly decidable:
An experiment can be devised (at least in principle) by which the truth values of all elementary propositions can be jointly ascertained.
Arbitrary repetitions of this experiment will reproduce with certainty the same result.
Such a most refined experiment yields as evidence a string of truth values $0$ or $1$.
The length of this string is a measure for the maximum amount of evidence that can be garnered from experiment.
It equals the cardinality of the hypothesis space.
At least in theory, therefore, the amount of evidence that an experimenter can accumulate matches the size of the hypothesis space.

In quantum theory the situation is radically different \cite{hartle:individual}.
There are propositions pertaining to non-commuting observables that are not jointly decidable.
For a finite-dimensional quantum system the total amount of reproducible evidence that can be garnered from experiment is bounded from above by the Hilbert space dimension and hence by a finite number;
whereas the hypothesis space comprises all possible pure states and hence is a continuous manifold.
The amount of evidence that any experimenter can accumulate is thus strictly \textit{smaller} than the hypothesis space, not due to practical limitations but as a matter of principle;
maximal information is not complete \cite{epr:complete,bohr:complete}.

It is the aim of this paper to show that not only does quantum theory imply a mismatch between hypothesis space and available evidence, but the converse is also true:
Whenever one is confronted with a situation where hypotheses form a continuum but total evidence is not allowed, as a matter of principle, to exceed a finite upper bound then plausible reasoning about that continuum of hypotheses, if it is to satisfy some basic consistency requirements, must necessarily follow the rules of quantum theory.

The proof of this claim presupposes that some basic consistency requirements for plausible reasoning---with the notable exception of ``joint decidability''---carry over from the classical case;
they will be detailed below.
As the principal subject of inquiry I will then introduce the group of ``consistency-preserving'' transformations in the continuous hypothesis space.
Analysis of this group, which to a good part amounts to a simple dimension-counting exercise, reveals that it must be isomorphic to the unitary group $U(d)$ where $d$ is the finite upper bound on the evidence.
This mandates the use of complex Hilbert space, and the identification of propositions with its subspaces, as the sole consistent framework for plausible reasoning.

Inferring the Hilbert space structure of quantum theory by means of a dimensional analysis has been proposed before \cite{hardy:fiveaxioms}.
Like the proof given below, that earlier proposal invoked the correspondence between probability distributions and measurements (state preparation), rules for the composition of systems, and the continuity of probabilities;
it focused on demonstrating that the manifold of (non-normalised) states has dimension $P(d)=d^2$.
However, it provided a rigorous proof only of $P(d)=d^\mu\;,\mu\in {\bf N}$, the cases $\mu\geq 3$ being excluded merely on the basis of a non-rigorous, albeit plausible, ``simplicity'' requirement.
In contrast to the approach presented here, the earlier proposal did not include a systematic study of the structure group and its dimension.
And finally, it made extensive use of the concepts ``pure state'' and ``fiducial state'', as well as of the language of linear vector spaces: notions that are inspired by quantum theory and already very suggestive of the structure to be derived, and that we will be trying to avoid here.

The remainder of this paper is organised as follows.
In Section \ref{basic} we introduce the basic notions of hypotheses, probabilities, filters and transformations.
The latter constitute a group, which will be the principal subject of our inquiry.
Section \ref{universality} provides a precise definition of ``maximum available evidence'', and argues that it alone determines the appropriate mathematical framework for plausible reasoning;
``evidence'' is the sole parameter of the theory.
Section \ref{analysis} constitutes the core of our analysis.
We inspect carefully, and formulate a number of consistency requirements pertaining to, the preparation and composition of systems, as well as the continuity of probabilities.
Thorough dimensional analysis then yields severe constraints on the structure group and leaves $U(d)$ as the only allowed choice.
We also discuss how this result may change if any of our assumptions are relaxed.
Finally, we wrap up our investigation with some concluding remarks in Section \ref{conclusions}.
There is an appendix in which we give some technical proofs omitted in the main text.

\section{\label{basic}Basic notions}

\subsection{\label{hypotheses}Hypotheses and probabilities}

We are concerned with hypotheses about some given physical system. 
Some (but not all) of these hypotheses may be related by logical implication: 
$x\subseteq a$ means that if hypothesis $x$ is true then hypothesis $a$ is also true;
hypothesis $x$ is a ``refinement'' of hypothesis $a$.
We shall denote the set of all possible refinements of a hypothesis $a$ by
\begin{equation}
	{\cal L}_a:=\left\{x\mid x\subseteq a\right\}\ .
\end{equation}
Logical implication constitutes a partial order: It is (i) reflexive, $x\subseteq x$; (ii) antisymmetric, $x\subseteq y,\; y\subseteq x \Rightarrow x=y$; and (iii) transitive, $x\subseteq y,\; y\subseteq z \Rightarrow {x\subseteq z}$.
There is a unique null element $\emptyset$, sometimes called the ``absurd hypothesis'', which is always false and hence implies all others (ex falso quodlibet): $x=\emptyset\Leftrightarrow x\subseteq a\;\forall a$.

A probability distribution $\rho$ assigns to each hypothesis a real number between $0$ and $1$.
We shall denote the set of all probability distributions on ${\cal L}_a$ by ${\cal P}_a$.
These two sets, ${\cal L}_a$ and ${\cal P}_a$, are dual to each other in the sense that any $\rho\in{\cal P}_a$ is completely specified by $\{\rho(x)\mid x\in{\cal L}_a\}$, and conversely any $x\in{\cal L}_a$ is completely specified by $\{\rho(x)\mid \rho\in{\cal P}_a\}$.
In ${\cal P}_a$ there is a partial order mirroring that in ${\cal L}_a$, defined by
\begin{equation}
	\rho\leq \sigma\ :\Leftrightarrow\ \rho(x)\leq \sigma(x)\ \forall x\in{\cal L}_a\ .
\end{equation}

In keeping with the Bayesian spirit we do not make any reference to limits of relative frequencies but only demand that the assignment of probabilities satisfy a number of consistency requirements.
Probability distributions must satisfy the common sense requirement that the more refined a hypothesis, the smaller its probability of being true; for any $x, y\in{\cal L}_a$,
\begin{equation}
	x\subseteq y\ \Leftrightarrow\ \rho(x)\leq \rho(y)\ \forall\rho\in{\cal P}_a\ .
\end{equation}
Probabilities are calibrated such that $\rho(\emptyset)=0$;
whereas they need not necessarily be normalised, i.e., the probability of the maximal element $a\in{\cal L}_a$ may be smaller than one.

When an observer assigns to a system either of the two probability distributions $\rho$ or $\sigma$ with respective ``probability of probabilities'' \cite{srednicki:probabilities} $\mbox{\rm prob}(\rho)$ and $\mbox{\rm prob}(\sigma)$ then, on this meta-level, the resulting probability for a hypothesis $x$ being true is given by the classical Bayes rule
\begin{eqnarray}
	\mbox{\rm prob}(x) &=& \mbox{\rm prob}(x\!\mid\!\rho)\cdot\mbox{\rm prob}(\rho)+\mbox{\rm prob}(x\!\mid\!\sigma)\cdot\mbox{\rm prob}(\sigma)\nonumber \\
	&=& \mbox{\rm prob}(x\mid\mbox{\rm prob}(\rho)\cdot\rho+\mbox{\rm prob}(\sigma)\cdot\sigma) \ ,
\end{eqnarray}
where $\mbox{\rm prob}(x\!\mid\!\rho)=\rho(x)$ (and likewise for $\sigma$).
Such mixing thus yields a new probability distribution which, being perfectly consistent, must also be contained in the set ${\cal P}_a$.
The latter is therefore convex:
\begin{equation}
	\rho,\sigma\in{\cal P}_a\ \Rightarrow\ t\rho+(1-t)\sigma\in{\cal P}_a\ \forall\;t\in [0,1]\ .
\end{equation}
Since we do not require probability distributions to be normalised, arbitrary rescaling is permitted as long as probabilities never become greater than one: 
\begin{equation}
	\rho\in{\cal P}_a\ \Rightarrow\ s\rho\in{\cal P}_a\ \forall\;s\in [0,1/\rho(a)]\ .
\end{equation}

\subsection{\label{filters}Filters}

There are further consistency requirements related to the processing of experimental evidence.
We imagine an experiment---we call it a ``filter''---that tests a certain hypothesis $b$ and then keeps the system only if $b$ is true, or else discards it if $b$ is false. 
In the course of such an experiment the experimenter will acquire new information and consequently update the probability distribution in two steps, which are to be carefully distinguished: 
(i) upon learning that the filter has been applied, yet with outcome still unknown; 
and (ii) upon learning about the outcome.
This can be summarized graphically as follows:
\begin{equation}
	\rho\ \stackrel{\rm(i)}{\rightarrow}\ \pi_b\rho\ \stackrel{\rm(ii)}{\rightarrow}\ \left\{
	\begin{array}{ll}
	\frac{1}{\rho(b)}\pi_b\rho & \mbox{if $b$ true} \\
	0                        & \mbox{otherwise (system discarded)}
	\end{array}
	\right.
\end{equation}
Step (i) introduces a---yet to be specified---map $\pi_b$ whose required properties will be discussed below; whereas step (ii) is a simple rescaling that carries over directly from the classical Bayes rule.

After step (i) all post-filter (but pre-reading) probabilities are bounded from above by the system's survival probability, 
\begin{equation}
	\pi_b\rho\:(x)\leq \rho(b)\ .
\end{equation}
In general these equal their prior values only for the hypothesis being tested and its refinements,
\begin{equation}
	\pi_b\rho\:(x)= \rho(x)\ \forall \rho\ \Leftrightarrow\  x\subseteq b\ .
\end{equation}
Filtering must preserve the partial order of probability distributions,
\begin{equation}
	\rho\leq \sigma\ \Leftrightarrow\ \pi_b\rho\leq \pi_b\sigma\ \forall b\ .
\end{equation}
Finally, two filters can be applied in series (without intermediate reading of results). If one of the hypotheses being tested is a refinement of the other then one may just as well apply the finer filter only; the coarser filter becomes redundant:
\begin{equation}
	b\subseteq a\ \Leftrightarrow\ \pi_b\circ\pi_a=\pi_a\circ\pi_b=\pi_b\ .
\end{equation}
However, for arbitrary hypotheses not related by logical implication the order in which the respective filters are applied may matter, so we do \textit{not} require that $\pi_b\circ\pi_a=\pi_a\circ\pi_b$ holds for every $a, b$.

Two hypotheses are said to ``contradict'' each other, \mbox{$a\perp b$}, if whenever one of them is true the other must be false.
Operationally this means that successive application of the respective filters must always lead to the system being discarded:
\begin{equation}
	a\perp b\ :\Leftrightarrow\ \pi_a\circ\pi_b=\pi_b\circ\pi_a=0\ .
\end{equation}
A set of hypotheses $\{b_i\}$ shall be called a ``set of alternative refinements'' of $a$ if they are mutually exclusive, $b_i\perp b_j\:\forall i\neq j$, while $b_i\subseteq a\:\forall i$; 
the set is ``complete'' if the refinements are also collectively exhaustive,
\[
	x\perp b_i\:\forall i\ \Leftrightarrow\ x\perp a\ .
\]
An incomplete set of alternative refinements can always be made complete by adding to it the unique hypothesis ``$a$, but not any of $\{b_i\}$''.
For a complete set of alternative refinements we require that the classical sum rule carry over,
\begin{equation}
	\{b_i\}\prec a\ \Leftrightarrow\ \rho(a)=\sum_i \rho(b_i)\;\forall\rho\ ,
\label{sumrule}
\end{equation}
where we have defined ``$\prec$'' as meaning that $\{b_i\}$ is a complete set of alternative refinements of $a$.

If a system is described by a mixture of two probability distributions $\rho,\sigma$ then application of the filter $\pi_b$ leads to a posterior probability 
\begin{eqnarray}
	\mbox{\rm prob}(x\!\mid\!\pi_b) &=& \mbox{\rm prob}(x\!\mid\!\rho,\pi_b)\cdot\mbox{\rm prob}(\rho\!\mid\!\pi_b)+\nonumber \\
	&&+\mbox{\rm prob}(x\!\mid\!\sigma,\pi_b)\cdot\mbox{\rm prob}(\sigma\!\mid\!\pi_b)  
\end{eqnarray}
for any $x\in{\cal L}_a$, where again on the meta-level we have invoked the classical Bayes rule. 
Requiring that the ``probability of probabilities'' is not affected by the presence or absence of the filter,
\begin{equation}
	\mbox{\rm prob}(\rho\!\mid\!\pi_b)=\mbox{\rm prob}(\rho)\ ,
\end{equation}
and using $\mbox{\rm prob}(x\!\mid\!\rho,\pi_b)=\pi_b\rho\:(x)$ we find that 
\begin{equation}
	\mbox{\rm prob}(x\!\mid\!\pi_b)=\mbox{\rm prob}(x\mid\mbox{\rm prob}(\rho)\cdot\pi_b\rho+\mbox{\rm prob}(\sigma)\cdot\pi_b\sigma)\ ,
\end{equation}
i.e., the map $\pi_b$ is linear:
\begin{equation}
	\pi_b(u\rho+v\sigma)=u\:\pi_b\rho+v\:\pi_b\sigma\ .
\end{equation}

\subsection{\label{transformations}Transformations}

Besides filtering, a second important class of experiments are ``transformations'' that do not involve the testing of any hypothesis, and whose effect amounts to a mere consistent relabeling of hypotheses; 
here ``consistency'' means that logical implications must be preserved.
Such consistency-preserving transformations form a group ${\cal G}_a$ of automorphisms of ${\cal P}_a$ and ${\cal L}_a$, respectively, that satisfy
\begin{equation}
	(g(\rho))(x)=\rho(g^{-1}(x))
\end{equation}
and
\begin{equation}
	x\subseteq y\ \Leftrightarrow\ g(x)\subseteq g(y)\ .
\end{equation}
The two types of experiment, (irreversible) filtering and (reversible) transformation, may be combined.
Their order of execution can be exchanged provided the hypothesis being tested is subjected to relabeling, too:
\begin{equation}
	g\circ \pi_b = \pi_{g(b)}\circ g\ .
\label{exchange}
\end{equation}

If a system is described by a mixture of two probability distributions $\rho,\sigma$ then, by a now familiar line of reasoning, transformation with $g\in{\cal G}_a$ leads to a posterior probability 
\begin{eqnarray}
	\mbox{\rm prob}(x\!\mid\!g) &=& \mbox{\rm prob}(x\!\mid\!\rho,g)\cdot\mbox{\rm prob}(\rho\!\mid\!g)+\nonumber \\
	&&+\mbox{\rm prob}(x\!\mid\!\sigma,g)\cdot\mbox{\rm prob}(\sigma\!\mid\!g)  
\end{eqnarray}
for any $x\in{\cal L}_a$. 
Requiring that the ``probability of probabilities'' is not affected by group action,
\begin{equation}
	\mbox{\rm prob}(\rho\!\mid\!g)=\mbox{\rm prob}(\rho)\ ,
\end{equation}
and using $\mbox{\rm prob}(x\!\mid\!\rho,g)=(g(\rho))\:(x)$ we find that 
\begin{equation}
	\mbox{\rm prob}(x\!\mid\!g)=\mbox{\rm prob}(x\mid\mbox{\rm prob}(\rho)\cdot g(\rho)+\mbox{\rm prob}(\sigma)\cdot g(\sigma))\ ,
\end{equation}
hence transformations are linear on ${\cal P}_a$:
\begin{equation}
	g(u\rho+v\sigma)=u\:g(\rho)+v\:g(\sigma)\ .
\end{equation}

\section{\label{universality}Evidence as the sole parameter}

Having ascertained the truth of a certain hypothesis $a$, the maximum amount of \textit{additional} evidence that can still be garnered from a most refined experiment equals the maximum number of alternative refinements of $a$; it shall be denoted by
\begin{equation}
	d(a):=\max\#\{b_i\mid \{b_i\}\prec a,\; b_i\neq\emptyset\}\ 
\end{equation}
and has the obvious properties
\begin{equation}
	x\subseteq y\ \Rightarrow\ d(x)\leq d(y)\ ,
\end{equation}
\begin{equation}
	d(x)=0\ \Leftrightarrow\ x=\emptyset\ .
\end{equation}
Furthermore, it is group-invariant,
\begin{equation}
	d(g(x))=d(x)\ \forall\ x\in{\cal L}_a,\ g\in{\cal G}_a\ .
\end{equation}
We are concerned with situations in which this maximum evidence is finite.

The above definition can be extended to probability distributions.
For any probability distribution $\rho\in{\cal P}_a$ we first define its ``support'' ${\rm supp}(\rho)$ as the unique hypothesis in ${\cal L}_a$ for which
\begin{equation}
	{\rm supp}(\rho)\subseteq x\ \Leftrightarrow\ \pi_x\rho=\rho\ .
\end{equation}
The support transforms in a covariant fashion,
\begin{equation}
	{\rm supp}(g(\rho))=g({\rm supp}(\rho))\ ,
\end{equation}
and after filtering is constrained to be a refinement of the hypothesis just verified,
\begin{equation}
	{\rm supp}(\pi_x\rho)\subseteq x\ ,
\end{equation}
with strict inequality if and only if $x$ has some non-absurd refinement whose probability vanishes:
\begin{equation}
	{\rm supp}(\pi_x\rho) \subset x\ \Leftrightarrow\ \exists\:y\subseteq x\:,\:y\neq\emptyset:\;\rho(y)=0\ .
\label{inequality}
\end{equation}
We shall then define
\begin{equation}
	d(\rho):=d({\rm supp}(\rho))\ .
\end{equation}
Like its counterpart for hypotheses it is group-invariant, and it has the analogous properties
\begin{equation}
	\rho\leq\sigma\ \Rightarrow\ d(\rho)\leq d(\sigma)\ ,
\end{equation}
\begin{equation}
	d(\rho)=0\ \Leftrightarrow\ \rho=0\ .
\end{equation}
Filtering generally produces new evidence and hence leads to a narrowing of probability distributions,
\begin{equation}
	d(\pi_x\rho)\leq d(\rho)\ ,
\label{narrowing}
\end{equation}
even though it is \textit{not} necessarily ${\rm supp}(\pi_x\rho)\subseteq {\rm supp}(\rho)$.
 
We require that the ``maximum available evidence'' be the only parameter of the theory.
This requirement has several important ramifications. 
To begin with, a hypothesis $a$ can be decomposed into ever more accurate alternative refinements in an iterative, tree-like fashion by first identifying some initial complete set of alternative refinements, then decomposing each of these refinements into a further complete set of alternative refinements, and so on until this process comes to a halt because hypotheses cannot be refined any further. 
The absence of other parameters implies that regardless of the precise path chosen to arrive at such a maximal decomposition, the total number of outermost branches at the end of the process must always be the same and equal to the maximum evidence $d(a)$;
which entails
\begin{equation}
		\{b_i\}\prec a\ \Rightarrow\ d(a)=\sum_i d(b_i)\ .
\end{equation}
Furthermore, whenever two hypotheses are at the same level of coarse-graining, $d(a)=d(b)$, then the corresponding substructures must be isomorphic: ${\cal L}_a\sim {\cal L}_b$ and ${\cal P}_a\sim {\cal P}_b$. 
The latter therefore form an equivalence class that depends on the maximum evidence only, and that we shall denote by ${\cal L}(d)$ and ${\cal P}(d)$, respectively.
Likewise the associated structure group, too, depends on the maximum evidence only and shall be denoted by ${\cal G}(d)$.

Finally, any hypothesis $x$ can have only one group-invariant property: its level of coarse-graining, $d(x)$.
As long as they are at the same level of coarse-graining, two hypotheses can always be transformed into one another by some consistent relabeling,
\begin{equation}
	d(x)=d(y)\ \Rightarrow\ \exists\:g\in{\cal G}(d):\ y=g(x)
\end{equation}
for any $x,y\in{\cal L}(d)$.
Thus the set of all hypotheses at the same level of coarse-graining $k$,
\begin{equation}
	{\cal M}_k(d):=\{x\in{\cal L}(d)\mid d(x)=k,\:k\leq d\}\ ,
\end{equation}
constitutes a homogeneous space on which ${\cal G}(d)$ acts transitively.
The stability group of any $y\in {\cal M}_k(d)$ equals the product of ${\cal G}(k)$ acting on its substructure ${\cal L}_y\sim {\cal L}(k)$, and ${\cal G}(d-k)$ acting on
\begin{equation}
	\{x\in{\cal L}(d)\mid x\perp y\}\sim {\cal L}(d-k)\ ;
\end{equation}
hence the set ${\cal M}_k(d)$ can be written as the quotient
\begin{equation}
	{\cal M}_k(d)\sim {\cal G}(d)/{\cal G}(k)\otimes {\cal G}(d-k)\ .
\label{quotient}
\end{equation}
This result can be generalised to complete sets of alternative refinements.
The set
\begin{equation}
	{\cal M}_{\{k_i\}}(d):=\{\:\{x_i\}\prec I_d\mid d(x_i)=k_i,\: \sum_i k_i=d\}\ ,
\end{equation}
where we have defined $I_d$ as the maximal element of ${\cal L}(d)$ with $d(I_d)=d$,
again constitutes a homogeneous space on which ${\cal G}(d)$ acts transitively.
The stability group of any $\{y_i\}\in {\cal M}_{\{k_i\}}(d)$ now equals the product of all ${\cal G}(k_i)$ acting on the respective substructures ${\cal L}_{y_i}\sim {\cal L}(k_i)$; 
hence
\begin{equation}
	{\cal M}_{\{k_i\}}(d)\sim {\cal G}(d)/\bigotimes_i {\cal G}(k_i)\ .
\label{quotient2}
\end{equation}

\section{\label{analysis}Dimensional analysis}

\subsection{\label{preliminaries}Preliminaries}

The set of probability distributions ${\cal P}(d)$, the structure group ${\cal G}(d)$ and the set of hypotheses ${\cal M}_k(d)$ may be discrete or continuous.
In case they are continuous we shall denote the dimensions of the respective manifolds by
\begin{equation}
	P(d):=\dim{\cal P}(d)\ ,
\end{equation}
\begin{equation}
	G(d):=\dim{\cal G}(d)\ ,
\end{equation}
\begin{equation}
	M_k(d):=\dim{\cal M}_k(d)\ ,
\end{equation}
where the quotient representation (\ref{quotient}) immediately implies the relation
\begin{equation}
	M_k(d)=G(d)-G(k)-G(d-k)\ .
\label{mg}
\end{equation}
In the trivial case $d=1$ there is only a single hypothesis, and any (non-normalised) probability distribution is uniquely specified by the probability of this single hypothesis being true.
Therefore,
\begin{equation}
	P(1)=1\ .
\end{equation}

Classically, ${\cal M}_k(d)$ is a discrete set, ${\cal G}(d)$ is an equally discrete permutation group, and any (non-normalised) probability distribution is determined by $d$ continuous parameters; hence
\begin{equation}
	P_{\rm cl}(d)=d\ ,\ G_{\rm cl}(d)=0\ ,\ M_{k\;{\rm cl}}(d)=0\ .
\end{equation}
In contrast, we are concerned here with situations in which hypotheses form a continuum.
In the following we shall argue that then the only other consistent solution is
\begin{equation}
	P(d)=d^2\ ,\ G(d)=d^2\ ,\ M_{k}(d)=2k(d-k)
\end{equation}
corresponding to ${\cal G}(d)\sim U(d)$.
This will involve closer inspection of, and some additional assumptions pertaining to, (i) the preparation and (ii) composition of systems, as well as (iii) the continuity of probabilities.

\subsection{\label{preparation}Preparation}

Any knowledge about a physical system, embodied in a probability distribution $\rho\in{\cal P}(d)$, is the result of a series of experiments or ``preparation procedures'' \cite{peres:book} applied to an initial state of total ignorance
\begin{equation}
	\rho^{(0)}(x):=d(x)/d\ \forall\ x\in{\cal L}(d)\ .
\end{equation}
This initial state of total ignorance is characterised by invariance under the structure group,
\begin{equation}
	g(\rho^{(0)})=\rho^{(0)}\ \forall\ g\in{\cal G}(d)\ ,
\end{equation}
in accordance with the ``principle of indifference'' \cite{jaynes:book}.

Preparation procedures may be arbitrary combinations of 
(i) testing sets of mutually exclusive hypotheses; 
(ii) keeping or discarding the system, with respective probabilities that may depend on the outcome of the test; 
and (iii) transformations.
In mathematical terms, for any $\rho\in{\cal P}(d)$ there exist sets of alternative refinements $\{b^{(\alpha)}_i\}$, sets of associated rescaling factors $\{\lambda^{(\alpha)}_i\}$ that reflect the respective probabilities of keeping or discarding the system, as well as transformations $\{g^{(\alpha)}\}$ such that
\begin{eqnarray}
	\rho &=& \ldots \circ g^{(\alpha)}\circ\left(\sum_i\lambda^{(\alpha)}_i\pi_{b^{(\alpha)}_i}\right)\circ\ldots \nonumber\\
	&& \ldots\circ g^{(1)}\circ\left(\sum_j\lambda^{(1)}_j\pi_{b^{(1)}_j}\right)\:\rho^{(0)}\ .
\end{eqnarray}
Using linearity and the exchange rule (\ref{exchange}) all transformations can be shifted to the right and absorbed in $\rho^{(0)}$, leaving behind only filters (pertaining to transformed sets $\{\tilde{b}\}$ of alternative refinements) and rescaling factors.
In particular, one can define a sequence of posterior probability distributions
\begin{equation}
	\rho^{(\alpha)}=\left(\sum_i\lambda^{(\alpha)}_i\pi_{\tilde{b}^{(\alpha)}_i}\right)\:\rho^{(\alpha-1)}\quad,\quad \alpha\geq 1
\label{sequence}
\end{equation}
after the $\alpha$-th preparation procedure, that eventually terminates to yield $\rho$.

The left-hand side of the above iteration is some point on the manifold ${\cal P}(d)$, hence specified by $P(d)$ real parameters.
The right-hand side, on the other hand, is uniquely specified by defining (i) the set $\{\tilde{b}_i\}$ of alternative refinements and (ii) for each refinement $\tilde{b}_j$, if ascertained, an associated posterior distribution in ${\cal P}_{\tilde{b}_j}$. 
Let $k_i:=d(\tilde{b}_i)$ and hence $\{\tilde{b}_i\}\in {\cal M}_{\{k_i\}}(d)$. 
Then due to the quotient representation (\ref{quotient2}) the set of alternative refinements is specified by
\begin{equation}
	\dim {\cal M}_{\{k_i\}}(d)=G(d)-\sum_i G(k_i)
\end{equation}
real parameters;
and a posterior distribution in ${\cal P}_{\tilde{b}_j}\sim {\cal P}(k_j)$ is specified by $P(k_j)$ real parameters.
Equating the total number of parameters on the left-hand side and on the right-hand side of the iteration equation then yields a first constraint on the dimensions:
\begin{equation}
	P(d)=G(d)-\sum_i G(k_i)+\sum_i P(k_i)\ ,\ \sum_i k_i=d\ .
\end{equation}
In the special case $k_i=1$ for all $i$ one obtains, using $P(1)=1$, 
\begin{equation}
	G(d)=P(d)+(G(1)-1)\cdot d\ .
\label{constraintpreparation}
\end{equation}

\subsection{\label{composition}Composition}

Let a system be composed of two subsystems with respective maximum evidence $d^{(1)}$, $d^{(2)}$ and complete sets of alternative refinements $\{x_i^{(1)}\}\prec I_{d^{(1)}}$, $\{x_j^{(2)}\}\prec I_{d^{(2)}}$.
Then the combined hypotheses $\{(x_i^{(1)},x_j^{(2)})\}$---meaning ``hypothesis $x_i^{(1)}$ pertaining to system $1$ \textit{and} hypothesis $x_j^{(2)}$ pertaining to system $2$''---constitute a complete set of alternative refinements in the composite system.
(Here the Boolean operation ``and'' is used in a perfectly classical sense since the two hypotheses refer to different subsystems and are thus jointly decidable; 
whereas for more general settings we carefully refrain from defining any of the conventional Boolean operations.)
If the hypotheses about the subsystems are ``most refined'' then so are the combined hypotheses about the composite system,
\begin{equation}
	d(x_i^{(1)})=d(x_j^{(2)})=1\ \Rightarrow\ d((x_i^{(1)},x_j^{(2)}))=1\ ;
\end{equation}
which implies that the maximum evidence about the composite system is the product $d^{(1)}\cdot d^{(2)}$.

Probability distributions for the two subsystems are specified by $P(d^{(1)})$ or $P(d^{(2)})$ real parameters, respectively.
This means that there is a set of $P(d^{(1)})$ (not necessarily mutually exclusive) hypotheses $\{b_i^{(1)}\}$, and likewise a set of $P(d^{(2)})$ hypotheses $\{b_j^{(2)}\}$, such that the probabilities for these selected hypotheses uniquely determine the full distribution in the respective subsystem.
Then for the composite system the full probability distribution is uniquely specified by the $P(d^{(1)})\cdot P(d^{(2)})$ combined hypotheses $\{(b_i^{(1)},b_j^{(2)})\}$;
i.e., $P(d^{(1)}d^{(2)})=P(d^{(1)}) P(d^{(2)})$.
Given $P(1)=1$ this yields a second constraint on the dimensions \cite{hardy:fiveaxioms}:
\begin{equation}
	P(d)=d^\mu\quad,\quad \mu\in {\bf N}\ .
\label{constraintcompositionstates}
\end{equation}

A similar line of reasoning can be applied to the composition of transformations.
Isolated transformations of the two subsystems are specified by $G(d^{(1)})$ or $G(d^{(2)})$ real parameters, respectively.
Hence, assuming the structure groups to be Lie groups, there are associated Lie algebras with ${G}(d^{(1)})$ generators $\{X_i^{(1)}\}$ and ${G}(d^{(2)})$ generators $\{X_j^{(2)}\}$, respectively.
Then for the composite system there must be a larger Lie algebra whose generators are isomorphic to the $G(d^{(1)})\cdot G(d^{(2)})$ pairs $\{(X_i^{(1)},X_j^{(2)})\}$.
This implies $G(d^{(1)}d^{(2)})=G(d^{(1)}) G(d^{(2)})$ and thus a third constraint on the dimensions:
\begin{equation}
	G(d)=0\quad {\rm or}\quad G(d)=d^\nu\quad,\quad \nu\in {\bf N}\ .
\label{constraintcompositiontransformations}
\end{equation}

\subsection{\label{continuity}Continuity}

We require that probabilities change under transformations in a continuous fashion, where ``continuity'' shall be defined as follows.
Assuming that the structure group ${\cal G}(d)$ is a Lie group and hence endowed with a group-invariant distance measure then it is possible to define, for any (infinitesimal) $\delta>0$, an (infinitesimal) neighborhood of the identity element $1_{\cal G}$
\begin{equation}
	{\cal G}_\delta (d):=\{g\in{\cal G}(d)\mid {\rm dist}(g,1_{\cal G})< \delta\}\ .
\end{equation}
Given a probability distribution $\rho\in{\cal P}(d)$, all refinements of its support have non-vanishing probabilities that are greater than or equal to
\begin{equation}
	\epsilon(\rho):=\min \{\rho(x)\mid x\subseteq {\rm supp}(\rho)\:,\:x\neq\emptyset\} >0\ .
\end{equation}
Now ``continuity'' means that probabilities that were initially greater than zero not suddenly jump to zero upon an infinitesimal transformation;
in more rigorous mathematical terms,
\begin{eqnarray}
	\forall\:\epsilon(\rho)>0\;\exists\:\delta>0:\: g(\rho)\:(x)>0 &\forall& x\subseteq{\rm supp}(\rho)\:, \:x\neq\emptyset\:,\nonumber \\
	&& g\in{\cal G}_\delta (d)\ .
\end{eqnarray}
By virtue of Eq. (\ref{inequality}) this is equivalent to requiring
\begin{equation}
	{\rm supp}\left[\pi_{{\rm supp}(\rho)} g(\rho)\right]={\rm supp}(\rho)\ \forall\ g\in{\cal G}_\delta (d)\ .
\label{defcontinuity}
\end{equation}
In the remainder of this section we shall always assume that we are in the infinitesimal neighborhood $g\in{\cal G}_\delta (d)$.

For further analysis we introduce an arbitrary auxiliary hypothesis $b$, 
\begin{equation}
	{\rm supp}(\rho)\subseteq b \subseteq I_d\ ,
\end{equation}
where the respective levels of coarse-graining $k:=d({\rm supp}(\rho))$, $l:=d(b)$ and $d\equiv d(I_d)$ satisfy
\begin{equation}
	k\leq l\leq d\ .
\end{equation}
Moreover, we define three additional auxiliary hypotheses $z$, $b\backslash z$ and $b^*_z$ as follows:
\begin{equation}
	z:={\rm supp}\left[\pi_b g(\rho)\right]\subseteq b\ ,
\end{equation}
\begin{equation}
	\{b\backslash z,z\}:\prec b
\end{equation}
and
\begin{equation}
	\{b^*_z,b\backslash z\}:\prec I_d\ .
\end{equation}
These definitions imply 
\begin{equation}
	z,\:g({\rm supp}(\rho))\subseteq b^*_z\ .
\label{gsupprho}
\end{equation}
Within the continuous region the associated levels of coarse-graining take the values
\begin{equation}
	d(z)=k\ ,\  d(b\backslash z)=l-k\ ,\  d(b^*_z)=d-l+k\ . 
\label{dimauxiliaries}
\end{equation}
The proofs of (\ref{gsupprho}) and (\ref{dimauxiliaries}) are given in the appendix.

As ${\rm supp}(\rho)$ and $z$ are both refinements of $b$ and have the same level of coarse-graining $k$, they are both elements of the set
\begin{equation}
	\{x\in{\cal L}_b\mid d(x)=k,\:k\leq l\}\sim {\cal M}_k(l)\ ;
\end{equation}
hence given $b$, the hypothesis $z$ is uniquely specified by $M_k(l)$ real parameters.
Likewise, $z$ and $g({\rm supp}(\rho))$ are both refinements of $b^*_z$, again at the same level of coarse-graining $k$, and thus elements of the set
\begin{equation}
	\{x\in{\cal L}_{b^*_z}\mid d(x)=k,\:k\leq (d-l+k)\}\sim {\cal M}_k(d-l+k)\ ;
\end{equation}
so given both $b$ and $z$, and hence $b^*_z$, the transformed support $g({\rm supp}(\rho))$ is uniquely specified by $M_k(d-l+k)$ real parameters.
Therefore the total number of parameters needed to specify $g({\rm supp}(\rho))$ is the sum $M_k(l)+M_k(d-l+k)$,
which must equal the number of parameters that would have been needed \textit{without} the above auxiliary construction:
\begin{equation}
	M_k(d)=M_k(l)+M_k(d-l+k)\ .
\end{equation}
In combination with Eq. (\ref{mg}) this implies the fourth and final constraint on the dimensions:
\begin{equation}
	G(d)={\frac{G(2)-2G(1)}{2}} d(d-1)+G(1)\: d\ .
\label{constraintcontinuity}
\end{equation}

\subsection{\label{summary}Summary}

The four constraints (\ref{constraintpreparation}), (\ref{constraintcompositionstates}), (\ref{constraintcompositiontransformations}) and (\ref{constraintcontinuity}) together with Eq. (\ref{mg}) permit only three solutions:
(i) the ``classical case'' in which hypotheses constitute a discrete set, the structure group is equally discrete, and any (non-normalised) probability distribution is determined by $d$ continuous parameters:
\begin{equation}
	P_{\rm cl}(d)=d\ ,\ G_{\rm cl}(d)=0\ ,\ M_{k\;{\rm cl}}(d)=0\ ;
\end{equation}
(ii) a case in which the set of hypotheses is still discrete and probability distributions are still determined by $d$ continuous parameters, yet there is a continuous group introducing non-trivial phases:
\begin{equation}
	P_{\rm sc}(d)=d\ ,\ G_{\rm sc}(d)=d\ ,\ M_{k\;{\rm sc}}(d)=0\ ,
\end{equation}
corresponding to ${\cal G}(d)\sim U(1)^{\otimes d}$; 
we may think of this as a ``semiclassical case''; and
(iii) the only allowed case in which hypotheses form a continuum:
\begin{equation}
	P_{\rm qu}(d)=d^2\ ,\ G_{\rm qu}(d)=d^2\ ,\ M_{k\;{\rm qu}}(d)=2k(d-k)\ .
\end{equation}
Given that ${\cal G}(d)$ must be a compact Lie group this leads to ${\cal G}(d)\sim U(d)$ \cite{barut:book}.
This last case proves our original conjecture:
Whenever hypotheses form a continuum but evidence is restricted to be finite, the \textit{only} consistent framework for plausible reasoning is the complex Hilbert space framework of quantum theory.

One may wonder what happens if any of the constraints are relaxed.
Table \ref{tab:constraints} gives an overview of our requirements for consistent reasoning, the associated dimensional constraints, and the additional cases allowed if a constraint is relaxed in isolation.
The requirements pertaining to preparation and to the composition of states are not instrumental in---but perfectly consistent with---deriving the dimensionality of the structure group;
without the preparation requirement, however, the connection is lost between the group dimension and the dimension of the state manifold.
If, instead, the requirement pertaining to the composition of transformations is relaxed then on purely dimensional grounds one additional structure group becomes possible: $SO(d)\otimes SO(d)$.
This new structure group leaves the dimensions of the various manifolds of hypotheses ${\cal M}_k(d)$ unchanged but changes their topology, e.g., ${\cal M}_1(2)$ becomes isomorphic to the surface of a torus rather than the surface of a sphere.
The physical significance of such topologies that are not simply connected remains elusive.
However, they might be in conflict with the requirement that the set of probability distributions be convex \cite{hardy:fiveaxioms}.
Finally, if the continuity requirement is relaxed in isolation then the dimensions of group and state manifold, while constrained to be equal, may be higher powers of $d$. 
Again it is not clear what the physical significance of such a behavior would be.

\begin{table*}
\caption{\label{tab:constraints}Overview of requirements for consistent reasoning, associated dimensional constraints, and additional cases allowed if a constraint is relaxed in isolation.}
\begin{ruledtabular}
\begin{tabular}{lccc}
 & &\multicolumn{2}{c}{Extra cases allowed if relaxed}\\
 Requirement&Implied dimensional constraint&Dimensions&Structure group\\ \hline
 Preparation&$G(d)=P(d)+(G(1)-1)\: d$&$P(d)=d^\mu\; , \; \mu\neq\nu$ &--- \\
 Composition (states)&$P(d)=d^\mu$ &---&--- \\
 Composition (transformations)&$G(d)=0, d^\nu$&$G(d)=d(d-1)$ &$SO(d)\otimes SO(d)$ \\
 Continuity&$G(d)={\frac{G(2)-2G(1)}{2}} d(d-1)+G(1)\: d$&$G(d)=P(d)=d^\mu\; ,\; \mu\geq 3$&many \\
\end{tabular}
\end{ruledtabular}
\end{table*}

\section{\label{conclusions}Conclusions}

We have considered the non-classical situation in which hypotheses form a continuum, whereas the maximum available evidence is bounded from above by some finite integer $d$. 
Employing the basic notions of hypotheses, probabilities, filters and transformations, and invoking a small number of consistency requirements pertaining to the preparation and composition of systems, as well as to the continuity of probabilities, we have shown that then the group of consistency-preserving transformations must be isomorphic to $U(d)$.
Our proof highlights the pivotal role played by the finite maximum evidence alias Hilbert space dimension $d$ as the sole parameter of the theory, confirming an earlier intuition by Fuchs \cite{fuchs:qmasinfo1}.

We have thus singled out complex Hilbert space as the \textit{only} consistent framework for plausible reasoning.
Quantum theory is indeed an ``island in theoryspace'' \cite{aaronson:island} distinguished by a high degree of internal consistency.
In particular, alternative models in real \cite{stueckelberg:real} or quaternionic \cite{finkelstein:quaternionic} Hilbert spaces that are allowed by traditional quantum logic \cite{jauch:book} (but that have already run into difficulties for other reasons such as the lack of a de Finetti representation \cite{caves:definettistates}) now seem very difficult to justify.
We also note that nowhere did we make reference to specific length or energy scales;
hence even though quantum phenomena are most prevalent in the microscopic world, there is nothing in the above line of argument that restricts it to that domain.

Once identified with quantum theory in complex Hilbert space, the various notions of statistical inference employed in this paper can be easily translated into the familiar language of conventional quantum theory;
these correspondences are summarised in Table \ref{translation}.
As is well known, quantum theory entails a number of counterintuitive features.
We recall a few, using the terminology of this article:
(i) 
The classical Boolean operations ``and'', ``or'' are not well defined for arbitrary hypotheses. 
Indeed, even though in certain special cases they are implicit in our above definitions of $\pi_b$, $\perp$ or $\prec$, we have avoided employing these notions in our line of argument.
(ii) 
Some pairs of hypotheses are not jointly decidable, making quantum theory inherently probabilistic. 
(Two hypotheses $x,y\in {\cal L}(d)$ are said to be jointly decidable if there is a complete set of alternative refinements $\{b_i\}_{i\in I}\prec I_d$ with subsets of the index set $I_x, I_y\subseteq I$ such that $\{b_i\}_{i\in I_x}\prec x$ and $\{b_i\}_{i\in I_y}\prec y$.)
(iii) 
It is not possible to assign to all hypotheses a preexisting truth value, i.e., to mimic quantum theory with a hidden-variables theory \cite{mermin:bell}.

Niels Bohr once remarked that physics in general, and quantum theory in particular, was to be regarded ``not so much as the study of something a priori given'' but rather as the development of ``methods for ordering and surveying human experience'' \cite{bohr:book}.
I hope this paper will have further corroborated the deep truth of this statement, provided we interpret ``ordering and surveying human experience'' as meaning ``consistent reasoning about hypotheses pertaining to the physical world''.

\begin{table*}
\caption{\label{translation}Correspondence between the terminology of statistical inference employed in this paper and the terminology of conventional quantum theory.}
\begin{ruledtabular}
\begin{tabular}{lclc}
 \multicolumn{2}{c}{Statistical inference}& \multicolumn{2}{c}{Quantum theory}\\
 Name&Symbol or relation&Name&Symbol or relation\\ \hline
 Hypothesis&$x$&Projector&$\hat{P}_x$, $x$ is subspace of Hilbert space \\
 Probability distribution&$\rho$ &Density matrix&$\hat{\rho}$ \\
 Probability&$\rho(x)$&Probability &${\rm tr}(\hat{\rho}\hat{P}_x)$ \\
 Logical implication&$x\subseteq y$&---&$\hat{P}_x\hat{P}_y=\hat{P}_y\hat{P}_x=\hat{P}_x$ \\
 Filter&$\pi_b \rho$ &---&$\hat{P}_b\hat{\rho}\hat{P}_b$ \\
 Contradiction&$x\perp y$ &Orthogonality&$\hat{P}_x\hat{P}_y=\hat{P}_y\hat{P}_x=0$ \\
 Complete set of alternative refinements&$\{b_i\}\prec a$ &Orthogonal decomposition&$\hat{P}_a=\sum_i \hat{P}_{b_i}$ \\
 Transformation&$g(\rho)$ &Unitary transformation&$\hat{U}\hat{\rho}\hat{U}^\dagger$ \\
 Level of coarse graining&$d(x)$ &Dimension of subspace&${\rm tr}(\hat{P}_x)$ \\
 Most refined hypothesis&$d(x)=1$ &1-dim. subspace (ray)&$\hat{P}_x=|\chi\rangle\langle\chi|$ \\
\end{tabular}
\end{ruledtabular}
\end{table*}

\begin{acknowledgments}
I thank Berndt M\"{u}ller for critical reading of the manuscript and valuable feedback.
\end{acknowledgments}

\appendix*

\section{}

\subsection{Proof of Eq. (\ref{gsupprho})}

That $z\subseteq b^*_z$ follows directly from their respective definitions.
The second logical implication in Eq. (\ref{gsupprho}) can be shown as follows.
It is
\begin{equation}
	\rho\left(g^{-1}(b)\backslash {\rm supp}\left(\pi_{g^{-1}(b)}\rho\right)\right)=0
\end{equation}
and hence
\begin{equation}
	\rho\left(g^{-1}(b)\backslash g^{-1}(z)\right)=0\ ,
\end{equation}
which implies
\begin{equation}
	g(\rho)\:(b\backslash z)=0
\end{equation}
and further
\begin{equation}
	g({\rm supp}(\rho))\perp b\backslash z\ .
\end{equation}
This yields
\begin{equation}
	g({\rm supp}(\rho)) \subseteq b^*_z\ ,
\end{equation}
Q.E.D.

\subsection{Proof of Eq. (\ref{dimauxiliaries})}

Eq. (\ref{narrowing}) and group invariance imply
\begin{equation}
	d(\pi_b g(\rho))\leq d(g(\rho)) = d(\rho)\ ;
\end{equation}
while $b\supseteq {\rm supp}(\rho)$ and the continuity condition (\ref{defcontinuity}) yield
\begin{equation}
	d(\pi_b g(\rho))\geq d(\pi_{{\rm supp}(\rho)} g(\rho)) = d(\rho)\ .
\end{equation}
Together these inequalities give
\begin{equation}
	d(\pi_b g(\rho)) = d(\rho)
\end{equation}
and hence $d(z)=k$, Q.E.D.


\end{document}